\documentstyle[12pt]{article}
\begin{document}
\def \beq{\begin{equation}}
\def \eeq{\end{equation}}
\rightline{DOE/ER/40561-221-INT95-17-06}
\rightline{EFI-95-50}
\rightline{hep-ex/9508011}
\vspace{0.5in}
\centerline{\bf DETECTION OF THE RF PULSE}
\centerline{\bf ASSOCIATED WITH COSMIC RAY AIR SHOWERS
\footnote{Appendix to proposal for the Auger Air Shower Array}}
\vspace{0.5in}
\centerline{\it Jonathan L. Rosner}
\centerline{\it Institute for Nuclear Theory}
\centerline{\it University of Washington, Seattle, WA 98195}
\bigskip
\centerline{and}
\bigskip
\centerline{\it Enrico Fermi Institute and Department of Physics}
\centerline{\it University of Chicago, Chicago, IL 60637
\footnote{Permanent address.}}
\bigskip

\centerline{\bf ABSTRACT}
\medskip
\begin{quote}
A project to detect the radio-frequency pulse associated with extensive
air showers of cosmic rays is described briefly.  Prototype work is
being performed at the CASA/MIA array in Utah, with the intention of
designing equipment that can be used in conjunction with the Auger
Giant Array proposal.
\end{quote}

\section{Motivation}

As a result of work in the 1960's and 1970's \cite{Allan}, some of which has
continued beyond then, it is recognized that air showers of energy 10$^{17}$ eV
are accompanied by radio-frequency pulses, whose polarization and frequency
spectrum suggest that they are due mainly to the separation of positive and
negative charges of the shower in the Earth's magnetic field.  The most
convincing data have been accumulated in the 50--100 MHz frequency range.
However, opinions have differed regarding the strength of the pulses, and
atmospheric and ionospheric effects have led to irreproducibility of results.
In particular, there may also be pulses associated with cosmic-ray-induced
atmospheric discharges \cite{atm}.

A study is being undertaken of the feasibility of equipping the Auger array
with the ability to detect such pulses.  It is possible that the higher energy
of the showers to which the array would be sensitive would change the
parameters of detection.  Before a design for large-scale RF pulse detection
can be produced, it is necessary to retrace some of the steps of the past 30
years by conclusively demonstrating the existence of the pulses for 10$^{17}$
eV showers, and by controlling or monitoring some of the factors which led to
their irreproducibility in the past.

In this note we describe the prototype activity at the CASA/MIA site, mention
related activities, and set forth some considerations regarding plans for the
Auger project.  More concrete plans for RF detection must await the outcome of
protoyping work at the CASA/MIA site.

\section{CASA/MIA Prototype setup}

In order to verify the claim \cite{Allan} that 10$^{17}$ eV showers are
accompanied by RF pulses with significant energy in the 50--100 MHz range, a
prototype detector is being set up at the CASA/MIA site in Dugway, Utah.  This
section describes the status of that effort.

\subsection{Large-event trigger}

A trigger based on the coincidence of several muon ``patches'' was set
to select ``large'' showers with a rate of several per hour. The MIA
Patch-Sum trigger was sent into a fan-in/fan-out.  From this, the signal was
put into a LeCroy 821 Discriminator. This produced a pulse of height $-0.8$ V
(on 50 Ohm output).  The width of the output pulse was set to 200 nsec. The
frequency of the output could be varied by adjusting the threshold on the
discriminator.  The observed rates as a function of this threshold
were as follows:

\begin{center}
\begin{tabular}{||c|c|c||} \hline
Threshold & Patches & Rate \\ \hline
  $-350$ mV & 6 & 45 Hz (highly variable) \\
  $-400$ mV & 7 & $3.7 \pm 0.25$ Hz \\
  $-450$ mV & 8 & $2.21 \pm 0.19$ Hz \\
  $-500$ mV & 9 & $1.23 \pm 0.10$ Hz \\
  $-560$ mV & 10 & $1.00 \pm 0.07$ Hz \\
  $-690$ mV & 12 & $0.52 \pm 0.03$ Hz \\
  $-750$ mV & 13 & $ 5.8 \pm 0.8 $ min$^{-1}$ \\
  $-810$ mV & 14 & $0.65 \pm 0.18$ min$^{-1}$ \\
  $-870$ mV & 15 & $ 2.0 \pm 0.8$ hr$^{-1}$ \\ \hline
\end{tabular}
\end{center}

The rate varied somewhat depending on how noisy any particular patch is. In
particular, below 6 patches, the rate was not observed to be very stable.
The electronics was initially set up at the 15 patch level.  This was estimated
to be $5 \times 10^{16}$ eV to $10^{17}$ eV, based on the rate at $10^{18}$ eV
of 1/km$^2$/day/sr.

The performance of this trigger was monitored during two runs (17985 and 17987)
on June 30, 1995. During a period in which 26 million CASA triggers had been
registered (estimated to be about 16 days assuming a 20 Hz rate), 240
``large-event triggers'' had been registered, or about 15 per day. The counters
were reset on the evening of 6/29/95.  During the subsequent 12.5 hours, 7
counts were registered between 7:12 pm MDT 6/29 and 7:47 am MDT 6/30,
consistent with the above rate. This is considerably less than the hoped-for
several per hour or the 2 per hour noted in the table above.

On the morning of 6/30/95, the large-event trigger level was reset from $-869$
mV to $-822$ mV.  At this level, it was checked that several counts per hour
would be registered. The large-event trigger was used to key a transceiver
which then broadcast a digitally recorded voice message which was picked up
remotely.

Times of receipt of the trigger message were measured to an accuracy of about 5
seconds. During remote receipt of trigger notifications, other tasks were being
performed, such as setting up of antenna, monitoring of RF backgrounds, and
testing of preamplifier.  Thus it is possible that some notifications were
missed.  For future studies it would be very helpful to record ``large-event
triggers'' with a time stamp and, if possible, a pulse voltage level to
determine by what amount a given threshold was exceeded.

A file of events with at least 15 muon patches was generated after the day's
runs:

\begin{center}
\begin{tabular}{c c c c} \hline
Run       &    Events    &    Live time &    Events/minute \\
          &              &    (minutes) &                  \\ \hline
17985     &     446      &      354     & $1.26 \pm 0.06$  \\
17987     &     495      &      354     & $1.40 \pm 0.06$  \\
Both      &     941      &      708     & $1.33 \pm 0.04$  \\ \hline
\end{tabular}
\end{center}

The times of events in this file which matched within 5 seconds of those found
using the remote-monitoring system mentioned were recorded. Of 58 potential
matches between large-event triggers and events in the above file, only 27
actual matches were found.  Although this correlation appears higher than
accidental, it is clear that many large-event triggers failed to match with
events in the actual data record.  Consequently, more effort is being devoted
to construction of an efficient large-event trigger.  It is notable that in
measuring HiRes - MIA coincidences for events of 10$^{17}$ eV, a rate of
somewhat less than 1 per hour was achieved \cite{coincs}.

\subsection{Monitoring of RF noise environment}

It was a concern that the RF noise of the local electronics and the presence of
an extensive lightning-protection array might dictate the placement of one or
more antennas outside the periphery of the array, or might make the site
unsuitable altogether.  The behavior of a single CASA board was investigated at
the University of Chicago. The various clock signals were detected at short
distances ($< 1$ m) from the board, but a much more intense set of harmonics of
78 kHz emanated from the switching power supplies.  These harmonics persisted
well above 100 MHz.  At 144--148 MHz, they overlapped, leading to intense
broad-band noise.

An initial survey of RF noise at the CASA site was performed. On the basis of
the results, which indicated some RF noise within the array, it was decided to
perform an initial follow-up survey sitting just outside the array.  The
original log-periodic antenna used to detect RF pulses at Chacaltaya in the
1960's and 1970's was obtained, tested for bandwidth, taken out to Utah, and
used in a follow-up study of RF noise in the 60--85 MHz frequency range.
Sources of most strong RF signals in this range appeared to be due to either
the receiver itself or to local TV stations. Spectrum analysis techniques may
be suitable for removing such monochromatic signals.

\subsection{Near-term plans}

It is proposed to monitor the RF noise and to detect pulses by mounting the
log-periodic antenna near the CASA central trailer site, just above the
lightning protection grid.  A digital storage scope will be used to register
several microseconds of RF data on a rolling basis. These data will then be
captured and inspected visually upon receipt of a large-event trigger.

The experiment will be repeated using successively greater amounts of
amplification and narrower band-pass filters once these become available. The
filters are being developed at the University of Chicago.  Once the large-event
trigger has been demonstrated to select events of 10$^{17}$ eV and above,
permanent digital recording of coincident pulses will be undertaken.

Still to be performed are experiments which seek to monitor RF pulses at lower
frequencies and at greater distances from the array. For these pulses, whose
strengths may be correlated with atmospheric electric fields, it is planned to
monitor such fields with the help of a field mill.

A spectrum analyzer will be used to make a broader survey of the RF noise in
various frequency ranges and may be of help in detecting potential sources of
interference to RF communications in the Auger project.

\section{Recent information on related activities}

\subsection{FORTE, BLACKBEARD, SNO, and other projects requiring digitizers}

Discussions with John Wilkerson at the University of Washington have been very
productive. Wilkerson was engaged in projects at Los Alamos with the acronyms
FORTE and BLACKBEARD whose aim was detection of electromagnetic pulses,
including those produced by cosmic-ray-induced electromagnetic discharges, with
frequency ranges in the 30--100 MHz range.  Many of the fast-digitization and
memory problems appear to be identical to those in the proposal for a protoype
pulse detector at CASA/MIA.  Time-frequency plots have been obtained by the
BLACKBEARD project which are exactly those one would hope to generate in a
survey at CASA/MIA.

Wilkerson has also encountered requirements similar to ours for digitization of
SNO data.  His estimate is that one can use Maxim MAX 100 A/D chips for less
than \$1K per channel, but that feeding their output into memory may well
amount to another \$1K per channel. Other references on digitizers have been
obtained \cite{Atiya,Bryman}. Discussions with Wilkerson will continue, and
further discussions with Dan Holden at Los Alamos are envisioned.

\subsection{Status of GHz detection}

David Wilkinson, who visited the University of Chicago during the spring of
1995, has promised to look into the power radiated at frequencies of several
GHz, where new opportunities exist associated with the availability of
low-noise receivers.  At latest report he had planned to complete the
relevant calculations during the summer of 1995.

\subsection{Other options}

Dispersion between arrival times of GPS signals on two different frequencies
may serve as a useful monitor of air shower activity.  The possibility of
correlation of large showers with such dispersion events will be investigated.

It may be possible at the CASA/MIA site to monitor commercial broadcast
signals in the 55 - 88 MHz range to detect momentary enhancements associated
with large showers, in the same sense that meteor showers produce such
enhancements.  Television Channels 3 and 6, for which no nearby stations
exist, offer one possibility.

\section{Considerations for Auger project}

At present we can only present a rough sketch of criteria for detection in the
50--100 MHz range. Data would be digitized at a 500 MHz rate at each station
and stored in a rolling manner, with at least 10 microseconds of data in the
pipeline at any moment.  Upon receipt of a trigger signaling the presence of a
``large'' shower ($> 10^{17}$ eV), these data would be merged into the rest of
the data stream at each station.

Per station, we estimate the following additional costs, in US dollars,
for RF pulse detection:

\begin{center}
\begin{tabular}{l l c} \hline
Two antennas and impedance transformers: & 200 & (a) \\
Mounting hardware:                       & 100 & (b) \\
Cables and connectors:                   & 200 & (c) \\
Preamps and lightning protection:        & 100 & (d) \\
Digitization and memory electronics:     &2000 & (e) \\ \hline
Total per station:                       &2600 & (f) \\ \hline
\end{tabular}
\end{center}

\noindent

\noindent
(a) Two commercial log-periodic TV antennas with commercial 4:1 baluns;
crossed polarizations.  Difference signal to be detected.

\noindent
(b) Highly dependent on other installations at site.  Antennas are
to be pointed vertically but optimum elevation not yet determined.

\noindent
(c) Antennas are mounted near central data acquisition site of each station.

\noindent
(d) Commercial GaAsFET preamps and gas discharge tubes.

\noindent
(e) Subject to prototype development experience.  Power requirements
not yet known.

\noindent
(f) The number of stations equipped with RF detection will not exceed 2000
per array, but could easily be fewer, depending on prototype experience.
\bigskip

The above estimate assumes that one can power the preamps and DAQ electronics
from the supply at each station without substantial added cost.  It also
assumes that the ``large-event trigger'' will be available at each station.
A further assumption is that the difference signal suffices to characterize
the pulse.  Additional preamplification and DAQ electronics may be required if
this is not so. A major consideration may be the acquisition of antennas robust
enough to withstand extreme weather (particularly wind) conditions.

For detection at frequencies above or below 50--100 MHz, the criteria are not
yet well enough developed to permit any cost estimate.

\section{Acknowledgments}

I thank Jim Cronin for inviting me to consider these questions for the Auger
project, Lucy Fortson for help in surveying RF noise generated by a CASA board,
Kevin Green for setting up the ``large-event trigger,'' Dick Gustafson for
discussions and for information about SSC equipment, Gerard Jendraszkiewicz and
Dave Smith for technical advice in the design of the ``alert module,'' Larry
Jones for supplying the original antennas used on Mount Chacaltaya and for
discussions, Brian Newport for logistical help at the CASA/MIA site and for
generating the file of events from runs 17985 and 17987, Rene Ong for
performing the initial RF survey work at the CASA/MIA site, Dave Peterson for
the loan of equipment used for monitoring RF at the CASA/MIA site, Leslie
Rosenberg for first interesting me in this problem, Fritz Toevs for technical
advice and laboratory space at the University of Washington, Augustine M. Urbas
for help with antenna measurements and filter design, John Wilkerson for
discussions regarding electronics and RF pulse detection, Dave Wilkinson for
stressing the importance of on-line monitoring of events during the initial RF
survey work, and Bob Williams for laboratory space.  This work was performed in
part during a visit to the Institute for Nuclear Theory at the University of
Washington, and was supported in part by the United States Department of Energy
under Grant No.~DE FG02 90ER40560.

\end{document}